\documentclass[12pt]{article}
\title{V\"axj\"o interpretation-2003: realism of contexts}
\author{Andrei Khrennikov\footnote{Supported in part by the EU Human
Potential Programme, contact HPRN--CT--2002--00279 (Network on
Quantum Probability and Applications) and Profile Math. Modelling
in Physics and Cogn. Sc. of V\"axj\"o University.}\\
International Center for Mathematical\\
Modeling in Physics and Cognitive Sciences,\\
MSI, University of V\"axj\"o, S-35195, Sweden\\
Email: Andrei.Khrennikov@msi.vxu.se}
\begin{document}
\maketitle

\begin{abstract}We present a new variant of the V\"axj\"o interpretation:
contextualistic statistical realistic. Basic ideas of 
the V\"axj\"o interpretation-2001 are essentially clarified. We also discuss applications 
to biology, psychology, sociology, economy,...
\end{abstract}

\bigskip

\bigskip

The first version of the {\it V\"axj\"o interpretation} of quantum mechanics was presented in [1], see also [2],
after the conference ``Quantum Theory: Reconsideration of foundations'', V\"axj\"o, June-2001, on the basis
of numerous exciting discussions with participants. I was really surprised, that in spite of the formal
acceptance of the official {\it Copenhagen interpretation,} many people (having top-qualification in quantum physics)
still have doubts of various kinds and many of them are still looking for a {\it realistic interpretation.}
This dream about {\it quantum realism} was the main stimulus for my attempt to present a new version of the
realistic interpretation of quantum mechanics. The main problem was to create such an interpretation
in which realism would coexist with a rather strange (people like to say ``nonclassical'') behaviour of quantum
probabilities -- {\it Born's rule and interference of probabilities.}\footnote{ This problem was well 
known to founders of quantum theory, see, for example, the correspondence
between A. Einstein and E. Schr\"odinger (see [3] for English translation and comments, see also [4]). 
E. Schr\"odinger did not like the Copenhagen interpretation; in particular, he created his 
cat just to demonstrate absurdness of this interpretation. Unfortunately, people practically forgot about this,
see [4] for detail. We also recall that Schr\"odinger's cat was just a modification of 
Einstein's example  ``involving a charge of gunpowder in a state of unstable chemical equilibrium'',
see [4] (letter of Einstein to Schr\"odinger, 8 August 1935, see [3], p.78). But neither Einstein
nor Schr\"odinger could combine a realistic ensemble model with quantum statistics. In particular, 
Schr\"odinger wrote to Einstein that he will accept the realistic statistical interpretation of quantum mechanics if the interference of probabilities
would be explained, see [4]. Consequently the Copenhagen interpretation 
preserved its status of the official quantum ideology until present time.} 

In 2000 it was demonstrated [5], see also [6]-[10], that Born's rule and interference of
probabilities can be (quite easily) 
derived in the realistic framework.\footnote{And this is one of advantages of my contextual statistical approach, 
cf. G. Mackey [11],  A. Lande [12], G. Ludwig [13], K. Kraus [14], see also P. Busch, M. Grabowski, P. Lahti [15], S. Gudder [16],
and A. Holevo [17].  In our approach everything is trivial: complex amplitudes are
constructed automatically on the basis of the formula of total probability with interference term, see [6]-[10].
Moreover, besides the ordinary complex Hilbert space representation,
there exists the hyperbolic one and mixed hyper-trigonometric, see [10]. I emphasize that the study of 
former approaches, especially investigations of A. Lande [12] and G. Mackey [11], were very important for me. However, the starting 
point were the books of P. Dirac [18] and  R. Feynman [19] in which they payed attention to the mystery 
of quantum interference of probabilities.} The only thing that should not be neglected is 
{\it contextuality of probabilities} -- dependence of probabilities on complexes of physical conditions (physical contexts).
My investigations were induced by interest to frequency probability theory, see R. von Mises [20], see also [21].
In the frequency approach probabilities directly depend on collectives (random sequences) which are
associated with concrete complexes of physical conditions. This probability theory is contextual
from the very beginning.

Since 2001 I organized a series of conferences on foundations of probability and quantum mechanics
\footnote{See http://www.msi.vxu.se/aktuellt/konferens/index.html and [22]--[24].} and through intensive discussions 
(and, in particular, hard critique of some my colleagues, see [25], [26]) my ideas on interpretations of quantum mechanics  
are now essentially clearer. 

First of all I understood the difference between my contextualism and Bohr's contextualism, see Remark 1.

Then I understood the difference between my contextualism and contextualism of operational (empirisist)
interpretation.

Another new issue is understanding of the role of special {\it reference observables} which are used
in a concrete model for probabilistic representation contexts (e.g., in classical and quantum physical
models we use the {\it position} and {\it momentum} observables).

Finally, it became clear that, in fact, I discussed [1], [2] not an interpretation of quantum mechanics,
but a model -- statistical and contextual  -- of physical reality. The corresponding interpretation of 
quantum mechanics is obtained automatically for those models in which statistical data can be represented
by complex amplitudes. This is the V\"axj\"o interpretation of quantum mechanics: {\it contextualistic
statistical  realistic interpretation.}

By starting not from the formalism of quantum mechanics (calculus of probabilities in a complex Hilbert space), but
from the general {\it contextual statistical  model of reality,} we get the possibility to apply contextual
statistical methods in many domains of science: {\it biology, psychology, sociology, economy,...} In some special cases
we can use even the quantum probabilistic formalism.  Such new applications of powerful mathematical 
methods developed in quantum theory can induce revolutionary changes in  many
sciences. But  for us the quantum formalism is not the starting point. We should start with the general V\"axj\"o model 
of reality (physical, biological, psychological, social,...) and then test statistical data to find an appropriative
mathematical formalism. In general there are no reasons to hope to obtain the
complex {\it quantum-like representation.} For example, there
might appear models in which data cannot be represented by complex amplitudes, but by hyperbolic ones. In this case
we should use the formalism of {\it hyperbolic quantum mechanics,} [10]. \footnote{In particular, our approach
implies that quantum mechanics is not complete.}

\medskip

\centerline{\bf 1. Realism of contexts} 

\medskip

We start with the basic definition:

{\bf Definition 1.} {\it A physical {\it context}  $C$ is  a complex of physical conditions.}

\medskip

In principle, the notion of context can be considered as a generalization of a widely used notion of 
{\it preparation procedure} 
[15]. I prefer to use contextualistic and not preparation terminology. By using the preparation terminology
we presuppose the presence of an experimenter preparing physical systems for a measurement. By using the 
contextualistic terminology we need not appeal to experimental preparations, experimenter should appear
only on the stage of a measurement. Moreover, context need not be macroscopic. Of course, there exist {\it experimental contexts} -- preparation
procedures. However, in general contexts are not coupled to preparation procedures. I consider contexts
as {\it elements of physical reality} which exist independently of experimenters\footnote{We use the notion
``elements of physical reality'' in common sense. There is no direct coupling with the EPR sufficient 
condition for values of physical observables to be elements of physical reality, see [27]. 
Moreover, in general the V\"axj\"o model need not contain physical systems. Thus even the formulation
of the question: ``Can values of observables be considered
as objective properties of physical systems?'' -- 
is in general meaningless. We shall come to the problem 
of reality of quantum observables as observables  on physical systems (i.e., classical or EPR reality)
in section 11. There we shall present the V\"axj\"o model  completed by physical systems.} This is 
the cornerstone of my contextualistic viewpoint to physics (quantum as well as classical):

\medskip

{\bf Contexts are elements of  reality}

\medskip

To construct a concrete model $M$ of reality, we should fix some set of contexts ${\cal C},$ see definition 2.

{\bf Remark 1.} (Copenhagen and V\"axj\"o contextualisms) {\small Bohr's interpretation of 
quantum mechanics   is considered as contextualistic, see [28] for detailed analysis. However, 
we should sharply distinguish two types of contextualism: Copenhagen and V\"axj\"o contextualisms. For N. Bohr ``context''
had the meaning ``context of a measurement''. For example, in his answer to the EPR challenge 
N. Bohr pointed out that position can be determined only in {\it context of position measurement.}
For me ``context'' has the meaning a complex of physical conditions. As was underlined, a context 
is an element of physical reality and it has no direct relation to measurements (or existence of experimenters 
at all).\footnote{We remark that so far we do not speak about an interpretation of quantum mechanics. 
We are presenting
an approach to modeling of physical reality. The quantum representation is possible only for some 
class of models, ${\cal M}_{\rm{quantum}}.$ The class ${\cal M}_{\rm{quantum}}$
is a very special subclass of the class of contextualistic statistical models.}
For example, there exist contextualistic statistical models which cannot be represented in a complex Hilbert space --
so called hyperbolic quantum-like models, [10].} Moreover, a Bohrian measurement context is always {\it macroscopic},
our context -- a complex of physical conditions -- need not be macroscopic. 

\medskip

\centerline{\bf 2. Observables} 

\medskip

Suppose that there is fixed a set of observables\footnote{We shall 
denote observables by Latin letters, $a,b,...,$ and their values by Greek letters, 
$\alpha, \beta,...$  }
${\cal O}$  such that any observable $a\in {\cal O}$ can be measured
under a complex of physical conditions $C$ for any $C\in {\cal C}.$ 

There can be in principle defined other observables on contexts ${\cal C}$
which do not belong to the system ${\cal O},$  but they will define another
contextual model of reality, see definition 2.

We remark that our general V\"axj\"o-representation of reality 
does not contain physical systems, cf. footnote 5.
At the moment we do not (and need not) consider observables
as observables on physical systems. It is only supposed that if a context $C$ is fixed then
for any instant of time $t$ we can perform a measurement of any observable  $a \in {\cal O}.$

We do not assume that all these observables can be measured simultaneously;
so they need not be compatible. The sets of observables $ {\cal O}$ and contexts ${\cal C}$ are coupled 
through

\medskip

{\bf Axiom 1:} {\it For  any observable 
$a \in {\cal O},$ there are well defined contexts $C_\alpha$
corresponding to $\alpha$-filtrations: if we perform a measurement of $a$ under 
the complex of physical conditions $C_\alpha,$ then we obtain the value $a=\alpha$ with 
probability 1. It is supposed that the set of contexts ${\cal C}$ contains filtration-contexts $C_\alpha$
for all observables $a\in {\cal O}.$}

\medskip

\centerline{\bf 3. Probabilistic representation of contexts} 

\medskip

{\bf Axiom 2:} {\it There are defined contextual probabilities ${\bf P}(a=\alpha/C)$ for any 
context $C \in {\cal C}$ and any observable $a \in {\it O}.$}

\medskip

At the moment we do not fix
a definition of  probability. Depending on a choice of probability theory we can obtain
different models. For any $C\in {\cal C},$ there is defined the set of probabilities:
$$
E({\cal O}, C)= \{ {\bf P}(a=\alpha/C): a \in {\cal O}\}
$$
We complete this probabilistic data by $C_\alpha$-contextual probabilities:
$$
D({\cal O}, C)= \{ {\bf P}(a=\alpha/C),...,
{\bf P}(a=\alpha/C_\beta), {\bf P}(b=\beta/C_\alpha),...: a,b,... \in {\cal O}\}
$$
(we remark that $D({\cal O}, C)$ does not contain the simultaneous probability distribution of 
observables ${\cal O}).$
Data $D({\cal O}, C)$ gives a probabilistic image of the context $C$ through the system of observables ${\cal O}.$
Probabilities ${\bf P}(a=\alpha/C_\beta),...$ play the role of {\it structural constants} of a model. 
We denote by the symbol ${\cal D}({\cal O}, {\cal C})$ the collection of probabilistic data
$D({\cal O}, C)$ for all contexts  $C\in {\cal C}.$ There is defined the map:
\begin{equation}
\label{MP}
\pi :{\cal C} \to {\cal D}({\cal O}, {\cal C}), \; \; \pi(C)= D({\cal O}, C).
\end{equation}
In general this map is not one-to-one. Thus the $\pi$-image of contextualistic reality is very rough:
{\it not all contexts can be distinguished with the aid of probabilistic data produced by the class 
of observables ${\cal O}.$ }

Mathematically such probabilistic data can be  represented in various ways. In some special cases
it is possible to represent  data by complex amplitudes.  A complex amplitude (wave function) $\phi\equiv
\phi_{D({\cal O}, C)}$ is constructed by using a formula of total probability with $\cos$-interference
term, see [6]-[10] for extended exposition.
In this way we obtain the probabilistic formalism of 
quantum mechanics. In other cases it is possible to represent data by hyperbolic amplitudes\footnote{Such amplitudes
are constructed by using a formula of total probability with $\cosh$-interference
term (``hyperbolic interference''), see [10].}
and we obtain the probabilistic formalism of ``hyperbolic quantum mechanics," [9], [10].

\medskip

\centerline{\bf 4. Contextualistic statistical model (V\"axj\"o model)} 

\medskip

{\bf Definition 2.} {\it A contextualistic  statistical model of reality is a triple
\begin{equation}
\label{VM}
M =({\cal C}, {\cal O}, {\cal D}({\cal O}, {\cal C}))
\end{equation}
where ${\cal C}$ is a set of contexts and ${\cal O}$ is a  set of observables
which satisfy to axioms 1,2, and ${\cal D}({\cal O}, {\cal C})$ is probabilistic data 
about contexts ${\cal C}$ obtained with the aid of observables ${\cal O}.$}

We call observables belonging the set ${\cal O}\equiv {\cal O}(M)$ {\it reference of observables.}
Inside of a model $M$  observables  belonging ${\cal O}$ give the only possible references
about a context $C\in {\cal C}.$

\medskip

\centerline{\bf 5. Realistic interpretation of reference observables}

\medskip

Our general model can (but, in principle,
need not) be completed by some interpretation of reference observables $a\in {\cal O}.$ 
By the V\"axj\"o interpretation reference observables are interpreted as {\it properties of contexts:} 

\medskip

``If an observation of $a$ under a complex of physical
conditions $C \in {\cal C}$ gives the result $a=\alpha,$  then this value is interpreted as 
the objective property of the context $C$ (at the moment of the observation).''

\medskip

As always, a model is not sensitive to interpretation. Therefore, instead of the realistic V\"axj\"o interpretation,
we might use the Bohrian measurement-contextualistic interpretation, see Remark 1. However, by assuming 
the reality of contexts it would be natural to assume also the reality of observables which are used for
the statistical representation of contexts.
Thus we use the realistic interpretation both for contexts and reference observables. This is 
V\"axj\"o realism.

\medskip

\centerline{\bf 6. On the role of reference observables}

\medskip

Reader has already paid attention that reference 
observables play the special role in our model. I interpret the set ${\cal O}$ as a family of observables
which represent some fixed class of properties of contexts belonging ${\cal C}.$ For example, such a family can be chosen 
by some class of cognitive systems ${\it Z}_{\rm{cogn}}$ -- ``observers'' -- which were interested only in 
the ${\cal O}$-properties of contexts ${\cal C}$ (and in the process of evolution  they
developed the ability to ``feel'' these and only these properties of contexts).
The latter does not mean that observables ${\cal O}$ are not realistic. I would like 
just to say that observers $\tau \in {\it Z}_{\rm{cogn}}$ use only  observables 
${\cal O}.$  

We remark again that
there can exist other properties of contexts  ${\cal C}$ which are not represented by observables 
${\cal O}.$ The same set of contexts ${\cal C}$ can be the basis of various models of contextual reality:
$M_i= ({\cal C}, {\cal O}_i, {\cal D}({\cal O}_i, {\cal C})), i=1,2,....$ For example, such models
can be created by various classes of  cognitive systems ${\it Z}_{\rm{cogn},i}.$ 

Moreover,  we may exclude the spiritual element from observables. By considering ``observation''
as ``feeling''of a context $C$ by some system $\tau$ we need not presuppose that $\tau$ is  a cognitive system.
Such a $\tau$  can be, e.g., a physical system (e.g. an electron) which ``feel'' a context $C$
(e.g., electromagnetic-context).

{\bf Remark 2.} (Number of reference observables) In both most important physical models -- in classical and quantum models -- 
the set ${\cal O}$ of reference observables consists of {\bf two observables:} {\it position and momentum.}
I think that this number ``two'' of reference observables plays the crucial role (at least in the quantum model).

\medskip

\centerline{\bf 7. V\"axj\"o model outside physics}

\medskip

Our contextual
statistical realistic models of reality can be used not only in physics, but in any domain
of natural and social sciences. Instead of complexes of physical conditions, we can consider
complexes of biological, social, economic,... conditions -- contexts -- as elements of reality.
Such elements of reality are represented by probabilistic data obtained with the aid of 
reference observables (biological, social, economic,...).

In the same way as in physics in some special cases
it is possible to encode such data by complex amplitudes. In this way we obtain 
representations of some biological, social, economic,.... models in complex Hilbert spaces. 
We call them {\it complex quantum-like models.} These models describe the usual $\cos$-interference 
of probabilities.

Thus, when we speak, e.g.,  about a quantum-like mental model, this has nothing to do 
with quantum mechanics for electrons, photons, ... contained in the brain, see [29] for detail. A 
quantum-like mental model is a contextualistic  probabilistic model of brain and nothing more, []. 
There were found (at least preliminary) experimental evidences that in psychology
there can be obtained quantum-like (i.e., represented by complex probability amplitudes) statistical data,
see [30]; such data also  can be generated by some games, [31] (which have been called
``quantum-like games'' in [31]). 

\medskip

\centerline{\bf 8. Choice of a probability model}

\medskip

As was mentioned, any V\"axj\"o model $M$ should be combined
on some concrete probabilistic model describing probabilistic data ${\cal D}({\cal O}, {\cal C}).$
Of course, the Kolmogorov measure-theoretical model dominates in modern
physics. However, this is not the only possible model for probability, see [21]. In particular, 
I strongly support using of the frequency model [20], [21]. 
Here we shall use this model to describe probabilistic data.
It does not mean that other models which are used in physics
cannot be combined with some V\"axj\"o models. Of course, such a combination is not straightforward, see
[8] on the use of the contextual extension of the Kolmogorov model. We now present the frequency probabilistic description
of data ${\cal D}({\cal O}, C)$ for some $C \in {\cal C}.$

\medskip

\centerline{\bf 9. Frequency description of probability distributions} 

\medskip

By taking into account Remark 2, we consider a set of reference
observables ${\cal O}= \{ a, b \}$ consisting of two observables $a$ and $b.$ 
We denotes the sets of values (``spectra'') of the reference observables by symbols $X_a$ and $X_b,$
respectively.

Let $C$ be some context. In a series of observations of $b$ (which can be infinite in a mathematical model)
we obtain a sequence of values of $b:$
\begin{equation}
\label{KOL1}
x\equiv x(b/C) = (x_1, x_2,..., x_N,...), \;\; x_j\in X_b.
\end{equation}
In a series of observations of $a$ we obtain a sequence of values of $a:$
\begin{equation}
\label{KOL2}
y\equiv y(a/C) = (y_1, y_2,..., y_N,...), \;\; y_j\in X_a.
\end{equation}
We suppose that the principle of statistical stabilization for relative frequencies 
holds true and the frequency probabilities 
are well defined:
\begin{equation}
\label{KOL3}
p^b(\beta) \equiv {\bf P}_x( b=\beta)= \lim_{N\to \infty} \nu_N(\beta; x), \;\; \beta \in X_b;
\end{equation}
\begin{equation}
\label{KOL3a}
p^a(\alpha) \equiv {\bf P}_y( a=\alpha)= \lim_{N\to \infty} \nu_N(\alpha; y), \;\; \alpha\in X_a.
\end{equation}
Here $\nu_N(\beta; x)$ and $ \nu_N(\alpha; y)$ are frequencies of observations of values
$b=\beta$ and $a=\alpha,$ respectively (under the complex of conditions $C).$

Let $C_{\alpha},  \alpha\in X_a,$  be contexts  corresponding 
to  $\alpha$-filtrations, see Axiom 1.
By observation of $b$ under the context $C_\alpha$ we obtain a sequence:
\begin{equation}
\label{KOL4}
x^{\alpha} \equiv x(b/C_\alpha) = (x_1, x_2,..., x_{N},...), \;\; x_j \in X_b.
\end{equation}
It is also assumed that for  sequences of observations  $x^{\alpha}, \alpha\in X_a,$ 
the principle of statistical stabilization for relative frequencies 
holds true and the frequency probabilities are well defined:
\begin{equation}
\label{KOL5}
p^{b/a}(\beta/\alpha) \equiv {\bf P}_{x^{\alpha}}(b=\beta)= \lim_{N \to \infty} \nu_{N}(\beta; x^{\alpha}), \;\;
\beta \in X_b.
\end{equation}
Here $\nu_N(\beta; x^\alpha), \alpha\in X_a,$  are frequencies of observations of value
$b=\beta$ under the complex of conditions $C_\alpha.$ We obtain
probability distributions:
\begin{equation}
\label{KKK4}
{\bf P}_x(\beta), \;\; {\bf P}_y (\alpha), \;
{\bf P}_{x^{\alpha}}(\beta),\;\;\alpha\in X_a, \beta \in X_b.
\end{equation}
We can repeat all previous considerations by changing $b/a$-conditioning to  $a/b$-conditioning. We consider
contexts $C_\beta, \beta \in X_b,$ corresponding to selections with respect to values of the observable $b$ and the 
corresponding collectives $y^{\beta}\equiv y(a/C_\beta)$ induced by   observations of $a$ in contexts $C_\beta.$
There can be defined probabilities $p^{a/b}(\alpha/\beta)\equiv {\bf P}_{y^{\beta}}(\alpha).$  Combining these 
data with data (\ref{KKK4}) we obtain
$$
D({\cal O}, C)= \{ p^a(\alpha), p^b(\beta), p^{b/a}(\beta/\alpha), p^{a/b}(\alpha/\beta): \alpha\in X_a, \beta \in X_b\}
$$

This data gives a statistical contextual image of reality based on reference observables
$a$ and $b.$ As was remarked, there exist various mathematical methods for encoding 
of data $D({\cal O}, C),$ e.g., in some cases by complex amplitudes -- complex quantum-like
representations.

\medskip

\centerline{\bf 10. Representation in a complex Hilbert space}

\medskip

Let $M$ be a contextualistic  statistical model such that ${\cal O}$ contains only two 
observables $a$ and $b.$ For any context $C\in {\cal C},$ by using statistical data $D(a,b, C)$
we can compute a quantity $\lambda(\beta/\alpha, C), \alpha \in X_a, \beta\in X_b,$ see [6]-[10].
This quantity was called in [6]-[10] a {\it measure of statistical disturbance} (of the $b$-observable
by the $a$-observations under the context $C).$ If
\begin{equation}
\label{EN}
\vert \lambda(\beta/\alpha, C)\vert \leq 1
\end{equation}
for all  $\alpha \in X_a, \beta\in X_b,$ then data $D({\cal O}, C)$ can be represented (by using the 
formula of total probability with interference term) by a complex amplitude $\phi_C$ or in the abstract framework
by an element the unit sphere $U_1$ of the complex Hilbert space $H.$ Denote the family of all contexts which satisfy
to (\ref{EN}) by the symbol ${\cal C}^{\rm{tr}}.$ We have the map:
\begin{equation}
\label{M1}
J : {\cal C}^{\rm{tr}} \to U_1
\end{equation}
We emphasize that $J$ is determined by the reference observables $a$ and $b.$ Thus (\ref{M1}) is 
a Hilbert space representation of contexts determined by these concrete reference observables. The map $J$
is not one to one. Thus by representing contexts by complex amplitudes we lose a lot of information 
about contexts.

The map (\ref{M1}) induces [8] a map:
\begin{equation}
\label{M2}
L: {\cal O} \to L (H),
\end{equation}
where $L (H)$ is the set of self-adjoint operators.
Probability distributions of operators $\hat{a}= L(a)$ and  $\hat{b}= L(b)$ 
(calculated by using quantum Hilbert space framework) in the state $\phi_C$ 
coincide with $ p^a(\alpha)$ and  $p^b(\beta).$

If for a context $C$ we find that 
\begin{equation}
\label{EN1}
\vert \lambda(\beta/\alpha, C)\vert \geq 1
\end{equation}
then $C$ can be represented by a hyperbolic amplitude

\medskip

\centerline{\bf 11. Systems, ensemble representation}

\medskip

 We now complete the contextualistic statistical model 
by considering systems $\omega$ (e.g., physical or cognitive, or social,..)
Systems are also {\bf elements of realty. }
In our model a context $C \in {\cal C}$ is represented  by an ensemble $S_C$ of systems which have
been interacted  with $C.$ For such systems we shall use notation:
$$
\omega \hookleftarrow C 
$$
The set of all (e.g., physical or cognitive, or social)
systems which are used to represent all contexts $C\in {\cal C}$ is denoted by the symbol
$\Omega\equiv \Omega({\cal C}).$
Thus we have a map:
\begin{equation}
\label{VMM}
C \to S_C=\{ \omega\in \Omega:  \omega \hookleftarrow C \}.
\end{equation}
This is the ensemble representation of contexts. We set 
$$
{\cal S}\equiv {\cal S}({\cal C})=\{S: S=S_C, C \in {\cal C}\}.
$$
The ensemble representation of contexts is given by the map (\ref{VMM})
$$
I: {\cal C} \to {\cal S}
$$
Reference observables ${\cal O}$ are now interpreted as observables on systems $\omega\in \Omega.$
In principle, we can interpret values of observables as {\it objective properties} of systems.
 Oppositely to the very common opinion, such models (with realistic observables)
can have nontrivial quantum-like representations (in complex and hyperbolic Hilbert spaces)
which are based on the formula of total probability with interference terms.

Probabilities are defined as ensemble probabilities, see [21].

{\bf Definition 3.} {\it The ensemble representation of a contextualistic  statistical model 
$M =({\cal C}, {\cal O}, {\cal D}({\cal O}, {\cal C}))$ is a triple
 \begin{equation}
 \label{VM1}
 S(M) =({\cal S}, {\cal O}, {\cal D}({\cal O}, {\cal C}))
 \end{equation}
 where ${\cal S}$ is a set of ensembles representing contexts ${\cal C}$, 
 ${\cal O}$ is a  set of observables, and ${\cal D}({\cal O}, {\cal C})$ is probabilistic data 
 about ensembles ${\cal S}$ obtained with the aid of observables ${\cal O}.$}
 
 \medskip
 
 {\bf 12. Algebraic structure on the set of reference observables}
 
 \medskip
 
 We do not assume the presence of any algebraic
 structure on ${\cal O}.$ Even if these observables take values in some set endowed with an algebraic structure, e.g., 
 in ${\bf R},$ we do not assume that this structure induces (in the standard way) the corresponding algebraic
 structure on ${\cal O}.$ If $a,b \in  {\cal O}$ and take values in ${\bf R}$ it does not imply that $d=a+b$ 
 is well defined as observable on every context $C\in {\cal C}.$ In the general contextual approach it is very clear
 why we cannot do this. If $a$ and $b$ are not compatible, then we cannot measure they simultaneously under a context $C$ 
 at the fixed instant of time and form $d=a+b.$ But a reader may say: 
 
 ``You use the realistic interpretation of the reference observables in a model $M.$
 Thus one can form the sum $d=a+b.$''
 
 a). By the realistic contextualistic interpretation,
 $a(t)$ and $b(t)$ are objective properties
 of a context $C$ at the instant of time $t.$ There is  defined $d(t)=a(t) + b(t)$.

 b). By the realistic interpretation of the model with systems,  $a(\omega)$ and $b(\omega)$ are objective properties
 of a system $\omega.$ There is defined $d(\omega)=a(\omega) + b(\omega)$.

 However,  this is the ontic or ``hidden sum'' and the representation (\ref{M2}) cannot be extended to 
 such sums. Quantum theory cannot say us anything about $d=a+b$ as pointwise observable.  Of course,
 we can define the sum of operators $\hat{d}= \hat{a}+\hat{b},$ but in general this operator would represent 
 not the ontic observable $d,$ but another observable $d_{\rm{quant}}.$ Observables $d$ and $d_{\rm{quant}}$
 can have different probability distributions! (see [8]). Nevertheless (and this seems to be crucial in using of quantum theory),
 averages of these observables coincide:
 \begin{equation}
 \label{SUM}
 <d_{\rm{quant}}>\equiv <\hat{d}>= <d>
 \end{equation}
 This is a consequence of linearity of both quantum (Hilbert space) and classical probabilistic averages
 and the coincidence of probability distributions of reference observables and they representatives in 
 the Hilbert space.

\medskip

\centerline{\bf 13. Realist and empirisist interpretations of quantum mechanics}

\medskip

We emphasize again that up to now we have not been considering an interpretation of quantum mechanics. There was proposed
a contextualistic statistical model of physical reality. Sometimes this model can be mathematically described
by using the formalism of classical mechanics, sometimes quantum, sometimes hyperbolic and so on. However,
it is useful to discuss relation of our model to models of physical reality corresponding to various interpretations
of quantum mechanics. Here we follow to P. Busch, M. Grabowski, P. J. Lahti [15], de Muynck, De Baere, and Martens [32]
and L. Ballentine [33].

{\bf 13.1. Empirisist interpretation.} In this interpretation the formalism of quantum mechanics
does not describe reality as such. It only serves to calculate probabilities (relative frequencies) of certain 
phenomena that can be interpreted as corresponding to the results of a quantum measurement. The probabilities are 
conditioned on certain procedures, to be interpreted as quantum mechanical preparation procedures. Thus, the wave function 
or density operator  can be interpreted as symbolizing a preparation procedure; in the same way a hermitian operator
describes symbolically a quantum mechanical measurement. Wave function and hermitian operator are not thought to correspond to 
something existing in microscopic reality. They are just labels of (macroscopic) instruments that can be found 
in the laboratory. QM is thought to describe only (cor)relations of preparation acts and measurement phenomena,
 It is also important for our further considerations to underline that
in an empirisist interpretation of QM the eigenvalues of the hermitian operator do not play a significant role,
because these eigenvalues do not correspond to properties of the microscopic object. The empirisist interpretation
has achieved great popularity because its antimetaphysical flavor: physics must be about observables only, and about
nothing else. Hence, in this interpretation neither the wave function nor the observable must be taken as 
a property of the microscopic object system.

{\bf 13.2. Realist interpretation.} By this interpretation values of physical observables are considered as 
objective properties -- properties of objects (physical systems).

{\bf 13.3. V\"axj\"o interpretation: realism of contexts.} In the V\"axj\"o approach quantum mechanics (as a physical 
theory) is a particular contextualistic statistical model of reality in which the probabilistic data 
${\cal D}({\cal O}, {\cal C})$ can be encoded by complex amplitudes. This point of view to quantum
formalism induces the V\"axj\"o interpretation of quantum mechanics. This is a {\it contextualistic statistical
realistic interpretation} of quantum mechanics. And V\"axj\"o realism is realism of contexts and reference 
observables.

The V\"axj\"o interpretation of quantum mechanics is quite close to the empirisist interpretation. 
The crucial difference is that by the V\"axj\"o interpretation quantum mechanics 
is about reality -- reality of contexts, and not about preparation and measurement 
procedures. Contexts exist independently of our measurement activity and values of reference observables
$a\in {\cal O}$ are objective properties of contexts. The space-scale does not play any role, because
the description of reality is purely probabilistic. Quantum probabilistic behaviour is a consequence of complementarity
of information for reference observables. 
Such complementarity of information can take place at microscopic as well as macroscopic scales and,
moreover,  not only in physics, but in any domain of natural and social sciences.

We remark that by considering context as an element of reality we eliminated the important difference
between realist and empirisist interpretations -- {\it the wave function is considered as a description of 
the result of preparation rather than as a symbolic representation of the preparation itself.} If a model
$M$ has a quantum(-like) complex representation then the wave function represents context -- a complex of physical
(or biological,...) conditions.

{\bf 13.4. V\"axj\"o interpretation: realism of contexts, systems and observables.} Let us now consider the completed 
 V\"axj\"o model which contains physical systems, contexts are represented by ensembles of systems. Physical 
 observables are considered as objective properties of systems. As well as for general contextualistic
 model,   quantum mechanics (as a physical 
theory) is about a rather special class of contexts ${\cal C}^{\rm{tr}}$ such the probabilistic data 
${\cal D}({\cal O}, {\cal C}^{\rm{tr}})$ can be encoded by complex amplitudes. The only difference is that probabilities are
defined as ensemble probabilities. This interpretation of quantum mechanics is very close to the well known
ensemble interpretation which was strongly supported by A. Einstein, see introduction; L. Ballentine called 
it the statistical interpretation, see [33]. A difference is that in our model we start with reality of contexts
which can be (but need not be) represented by ensembles. 

But this is not the main difference. The main difference
is that we did not start at all with an interpretation of one special mathematical formalism, 
calculus of probabilities in complex Hilbert
spaces. We started with a general contextual statistical model of reality and 
then demonstrated that some special contexts
can be represented by quantum-like complex amplitudes. Interpretation of such amplitudes follows automatically
from the basic contextual statistical model.

\bigskip

{\bf References}

1. A. Yu. Khrennikov, V\"axj\"o interpretation of quantum mechanics, 
http://xxx.lanl.gov/abs/quant-ph/0202107.

2. A. Yu. Khrennikov, On foundations of quantum theory. 
Proc. Conf. {\it Quantum Theory: Reconsideration
of Foundations,} ed. A. Yu. Khrennikov. 
Ser. Math. Modelling, {\bf 2}, 163-196,V\"axj\"o Univ. Press (2002).

3. A. Fine, {\it The Shaky game.} Univ. Chicago Press, Chicago/London (1988).

4. M. Lockwood, What Schr\"odinger should have learned from his cat?
In {\it Erwin Schr\"odinger: Philosophy and the birth of quantum mechanics,} eds. M. Bitbol, O. Darrigol.
Editions Frontieres, Gif-sur Yvette (1992).'

5. A. Yu. Khrennikov, Origin of quantum probabilities. Proc. Conf. {\it Foundations of Probability
and Physics.} In: {\it Q. Probability and White Noise Analysis},
{\bf 13}, 180-200, WSP, Singapore (2001).

6. A. Yu. Khrennikov, {\it Linear representations of probabilistic
transformations induced by context transitions.} {\it J. Phys. A: Math. Gen.,} 
{\bf 34}, 9965-9981 (2001); http://xxx.lanl.gov/abs/quant-ph/0105059 
 
7. A. Yu. Khrennikov, Contextual viewpoint to quantum stochastics. 
{\it J. Math. Phys.}, {\bf 44},  2471- 2478 (2003).

8. A. Yu. Khrennikov, Representation of the Kolmogorov model having all distinguishing 
features of quantum probabilistic model. {\it Phys. Lett. A}, {\bf 316}, 279-296 (2003).

9.   A. Yu. Khrennikov, Hyperbolic quantum mechanics. {\it Advances in Applied Clifford Algebras,}
{\bf 13}(1),  1-9 (2003).
 
10. A. Yu. Khrennikov, Interference of probabilities and number field structure of quantum models.
{\it Annalen  der Physik,} {\bf 12},  575-585 (2003).

11. G. W. Mackey, {\it Mathematical foundations of quantum mechanics.}
W. A. Benjamin INc, New York (1963).

12. A. Lande, {\it Foundations of quantum theory.} Yale Univ. Press (1955).

A. Lande, {\it New foundations of quantum mechanics} Cambridge Univ. Press, Cambridge (1968).

13. G. Ludwig, {\it  Foundations of quantum mechanics.} Springer, 
Berlin (1983).

14. K. Kraus, {\it States, effects and operations.} Springer-Verlag, Berlin (1983).

15. P. Busch, M. Grabowski, P. Lahti, Operational Quantum Physics,
Springer Verlag (1995).

16. S. P. Gudder, {\it Axiomatic quantum mechanics and generalized probability theory.}
Academic Press, New York (1970).

17. A. S. Holevo, {\it Probabilistic and statistical aspects of quantum 
theory.} North-Holland, Amsterdam (1982).

A. S. Holevo, {\it Statistical structure of quantum theory.} Springer,
Berlin-Heidelberg (2001).

18. P. A. M.  Dirac, {\it The Principles of Quantum Mechanics.}
Oxford Univ. Press (1930).

19. R. Feynman and A. Hibbs, {\it Quantum Mechanics and Path Integrals.}
McGraw-Hill, New-York (1965).

20. R. Von Mises,  {\it The mathematical theory of probability and
 statistics.} Academic, London (1964).

21. A. Yu. Khrennikov, {\it Interpretations of Probability.}
VSP Int. Sc. Publishers, Utrecht/Tokyo (1999).

22. A. Yu. Khrennikov, ed., Proc. Int. Conf. {\it Quantum Theory: Reconsideration
of Foundations.} Ser. Math. Modelling, {\bf 2,}
 V\"axj\"o Univ. Press (2002).
 
23.   A. Yu. Khrennikov, ed., Proc. Conf.
{\it Foundations of Probability and Physics-2,} 
Ser. Math. Modelling,  {\bf 5,}
V\"axj\"o Univ. Press (2002).

24.  A. Yu. Khrennikov, ed., Proc. Int. Conf. {\it Quantum Theory: Reconsideration
of Foundations-2.} Ser. Math. Modelling, {\bf 10,}
 V\"axj\"o Univ. Press (2004).

25. C. Fuchs, The anti-V\"axj\"o interpretation of quantum mechanics.
Proc. Int. Conf. {\it Quantum Theory: Reconsideration
of Foundations.} ed. A. Yu. Khrennikov, Ser. Math. Modelling, {\bf 2},
99-116, V\"axj\"o Univ. Press,  2002;  http://www.msi.vxu.se/forskn/quantum.pdf

26. A. Plotnitsky, The spirit and the letter of Copenhagen: a response
to Andrei Khrennikov, http://xxx.lanl.gov/abs/quant-ph/0206026.

27.  A. Einstein, B. Podolsky, N. Rosen,  {\it Phys. Rev.}, {\bf 47}, 777--780
(1935).

28. A. Plotnitsky, Quantum atomicity and
quantum information: Bohr, Heisenberg, and quantum mechanics as an
information theory, Proc. Conf. {\it Quantum theory:
reconsideration of foundations,} ed: A. Yu. Khrennikov, Ser. Math. Modelling, 
{\bf 2}, 309-343, V\"axj\"o Univ. Press (2002).

A. Plotnitsky, Reading Bohr:
Complementarity, Epistemology, Entanglement, and Decoherence.
Proc. NATO Workshop ''Decoherence and its Implications for Quantum
Computations'', Eds. A.Gonis and P.Turchi, p.3--37, IOS Press,
Amsterdam, 2001.

29. A. Yu. Khrennikov, On cognitive experiments to test quantum-like behaviour 
of mind. {\it Rep. V\"axj\"o Univ.: Math. Nat. Sc. Tech.,} N 7 (2002);
http://xxx.lanl.gov/abs/quant-ph/0205092.

 A. Yu. Khrennikov, Quantum-like formalism for cognitive measurements. {\it Biosystems,}
{\bf 70}, 211-233 (2003).

30. E. Conte, O. Todarello, A. Federici, T. Vitiello, M. Lopane, A. Yu. Khrennikov 
A Preliminar Evidence of Quantum Like Behavior in Measurements of Mental States. 
{\it Reports from MSI, V\"axj\"o Univ.,} N. 03090, 2003;\\http://xxx.lanl.gov/abs/quant-ph/0307201. 

31. A. Grib, A. Khrennikov, K.Starkov, Probability amplitude in quantum like games.
{\it Reports from MSI, V\"axj\"o Univ.,} N. 03088, 2003;\\ http://xxx.lanl.gov/abs/quant-ph/0308074.

32. W. M. de Muynck, W. De Baere, and H. Martens, Interpretations of quantum mechanics, joint
measurement of incompatible observables, counterfactual definiteness. {\it Found. Physics,}
{\bf 24}, N. 12 (1994).

33. L. E. Ballentine,  {\it Rev. Mod. Phys.}, {\bf 42}, 358 (1970).

L. E. Ballentine, {\it Quantum mechanics.} Englewood Cliffs, 
New Jersey (1989).

\end{document}